\documentclass{aa}
\begin{document}

\title {Space-VLBI phase-reference mapping and astrometry}

\author{J.C.~Guirado\inst{1,2}
\and E.~Ros\inst{3} 
\and D.L.~Jones\inst{4}
\and J.-F.~Lestrade\inst{5}
\and J.M.~Marcaide\inst{1}
\and M.A.~P\'erez-Torres\inst{6}
\and R.A.~Preston\inst{4}
}

\offprints{J.C.~Guirado}

\institute{
Departamento de Astronom\'{\i}a, Universitat de Val\`{e}ncia, E-46100 Burjassot, Valencia, Spain
\and
Observatorio Astron\'omico, Universitat de Val\`{e}ncia, E-46100 Burjassot, Valencia, Spain
\and
Max-Planck-Institut f\"ur Radioastronomie, Auf dem H\"ugel 69, 
D-53121, Bonn, Germany 
\and
Jet Propulsion Laboratory, California Institute of Technology,
Pasadena, California 91109, USA
\and
Observatoire de Paris-Meudon-CNRS, F-92195 Meudon Principal Cedex,
 France
\and
Istituto di Radioastronomia, CNR, Via P. Gobetti, 101, I-40129 Bologna, 
Italy
}

\date{ } 

\abstract{
We present 5\,GHz space-VLBI observations of the quasar pair
B\,1342+662 / B\,1342+663 that demonstrate the feasibility of 
phase-reference techniques using an antenna in space. 
The space-based data were taken by the satellite HALCA, 
of the space-VLBI mission VSOP.
From residual (referenced) phases we derive
an upper bound of 10 meters to the uncertainty
of the spacecraft orbit reconstruction. An analysis
of the phase-reference maps of the sources
additionally suggests that the above mentioned uncertainty
is likely not larger than 3 meters.
With errors of this magnitude, HALCA is
a useful tool for astrometric studies of close pairs of
radio sources.
}

\maketitle

\section{Introduction}

The relativistic jet theory for compact extragalactic radio sources assumes 
that most of the 
emission from quasars and AGNs comes from a central engine, predicted
to be stationary at the microarcsecond ($\mu$as) level for standard
cosmologies, ignoring opacity effects. In spite of the high precision 
provided by techniques such as Very-Long-Baseline Interferometry (VLBI) 
differential astrometry, the experimental status of
this prediction is not satisfactory since only 3C\,345 (Bartel et al. 1986)
and 1038+528\,A/B (Marcaide et al. 1994) have been observed astrometrically
with high enough precision to verify this stationarity. 
One of the largest contributors to the error budget of the observations 
above was the uncertainty inherent in the
limited resolution of the radio source maps. The establishment of a global 
reference
frame has a similar limitation: no matter how good the astrometric model may be,
the lack of a reproducible point in the structure of the radio source, that may serve 
as reference, is the limiting factor of the precision. The latter 
is a necessary consequence of the 
nature of the observations: ground-based astrometry combines 
nearly {\it microarcsecond}-precise relative positions with only
{\it milliarcsecond}-size features of the source structure. 

Attempts to alleviate this situation should be directed to
the extension of the VLBI phase-reference mapping and astrometry techniques 
to observations at the highest possible resolution, i.e.,
mm-VLBI and space-VLBI. The feasibility of differential
astrometry at mm-wavelengths has already been demonstrated for a
pair of sources with 5\degr\, separation and high antenna
elevations (Guirado et al. 2000).
The launch in February 1997 of the HALCA (Highly Advanced Laboratory For 
Communication and Astronomy) satellite of the VLBI Space Observatory 
Program (VSOP; Hirabayashi et al. 1998) opened the study of 
radio sources at the highest possible resolution at cm-wavelengths. The 
potential capabilities 
of VSOP for astrometry are considerable, but, in practice, they are limited 
by the 
on-board command memory of the spacecraft and the uncertainties of the orbit 
reconstruction (Ulvestad 1999). However, an immediate advantage of space-VLBI 
observations is the improvement in the resolution of
the image that makes the selection of a reference point rather straightforward and less
affected by uncertainties.\\ 

In this paper, we explore the capabilities of VSOP for phase-reference 
mapping and astrometry by means of observations of the pair B\,1342+662/B\,1342+663 
(Morabito 1984). This quasar pair is extremely favorable for the purpose: the
separation of the sources is small enough so that both sources lie 
within the primary beam of the HALCA and the VLBA antennas. This particular 
configuration permits an advantageous use of the phase-reference technique, 
where the interpolation of the observables in time, needed by switching 
schemes, can be avoided (Marcaide \& Shapiro 1983). Systematic errors 
(among them, orbit reconstruction errors) are thus largely cancelled.
From the analysis of the phase-reference maps of the observed sources, 
upper bounds to the uncertainty of HALCA's orbit reconstruction can also 
be derived.

\section{Observations and Data Reduction}

\begin{figure}[htpb]
\vspace*{250pt}
\includegraphics{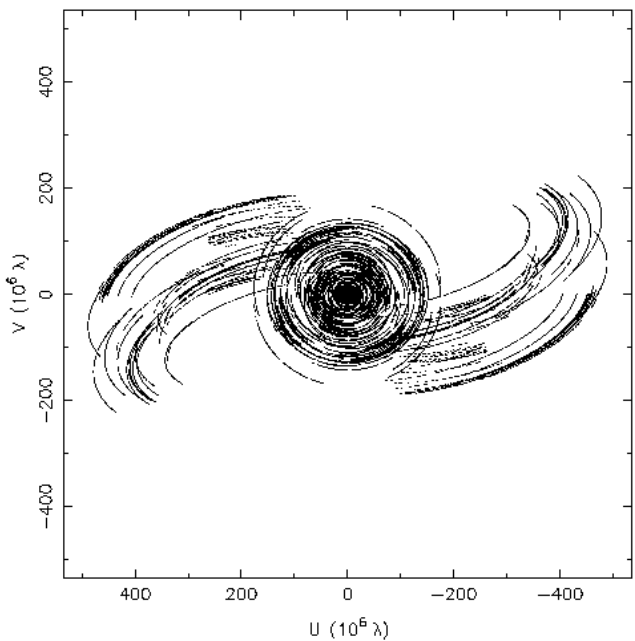}
\caption[]{  {\it uv}-coverage of our space-VLBI observations. The 
denser set of points with uv-distances less than 200\,M$\lambda$ 
corresponds to ground-based baselines. The outer tracks correspond 
to baselines to HALCA.} 
\end{figure}

We made observations of the radio sources B\,1342+662 and
B\,1342+663 on 1999 May 8 from 12:00 to 08:00\,UT (next day). We used 
the complete
VLBA (ten 25\,m antennas), Effelsberg (100\,m, Germany), and
Kashima (34\,m, Japan), along with
the Japanese satellite HALCA (8\,m) at a frequency of 4.8\,GHz, 
recording two intermediates frequency (IF) channels of 16\,MHz each, 
with a data rate of 128\,Mb/s. 
Five tracking stations received
the astronomical data from HALCA through a Ku-band downlink, 
namely, Usuda (Japan), Robledo (Spain), Green Bank (USA), 
Tidbinbilla (Australia),
and Goldstone (USA). We show in Fig. 1 the uv-coverage of our 
observations to emphasize the improvement in resolution provided by 
the baselines to HALCA. 

\begin{figure}[htpb]
\vspace*{400pt}
\includegraphics{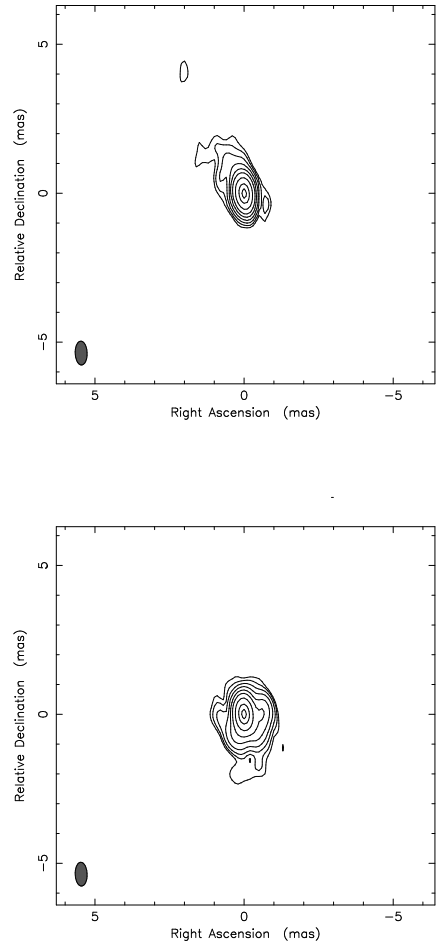}
\caption[]{5\,GHz hybrid maps of B\,1342+662 (upper panel) and B\,1342+663
(lower panel) at epoch 1999.35. Contours are 
-0.5,0.5,1,2,4,8,16,32,64, and 90\% of
the peak of brightness, 
0.21 Jy/beam (B\,1342+662), and 0.42 Jy/beam (B\,1342+663). 
The restoring beam (shown
at the bottom left corner of each map) is an elliptical Gaussian of
0.8$\times$0.4\,mas (P.A. 2\degr).}
\end{figure}

Since the separation of B\,1342+662 and B\,1342+663 is only 
0\fdg 08 ($4\farcm 8$) on the sky, both sources were observed simultaneously by 
HALCA and the ground telescopes, except Effelsberg (whose antenna beam 
at 4.8\,GHz being $2\rlap{.}'5$, switched between the sources; {however, 
the switching time could not be well optimized for astrometric purposes and 
the data of baselines to Effelsberg had to be finally discarded}). 
The data were correlated at the National Radio Astronomy Observatory 
(NRAO, Socorro, NM, USA). Two correlator passes were made, one at 
the position of each of the two sources.
For each source, and using the NRAO 
Astronomical Image Processing System ({\sc aips}), 
we made an a priori visibility amplitude
calibration (using system temperatures and gain curves from each antenna),
and a fringe fitting around the sky position used at correlation time. 
We optimized 
the 3\,min-interval fringe solutions with the following iterative procedure:  
first, for each IF, we found fringe solutions using jointly delay and delay 
rates; second, using the previous solution as a priori and for each 
IF, we repeated 
the fringe fitting with narrow search windows in delay (50 ns) and 
delay rate (100 mHz); finally, using the previous solution as a priori, 
we made a final fringe fitting for both IFs simultaneously to determine
single-band delay, multi-band delay, and delay rate.
The procedure described above was needed to overcome the lack of
phase-calibration information for the satellite data, that prevented 
detections in several baselines to HALCA with a standard fringe fitting.
As a result of our fringe fitting procedure, both sources were detected in 
all baselines, including those to HALCA.

\begin{figure*}[htpb]
\vspace*{230pt}
\includegraphics{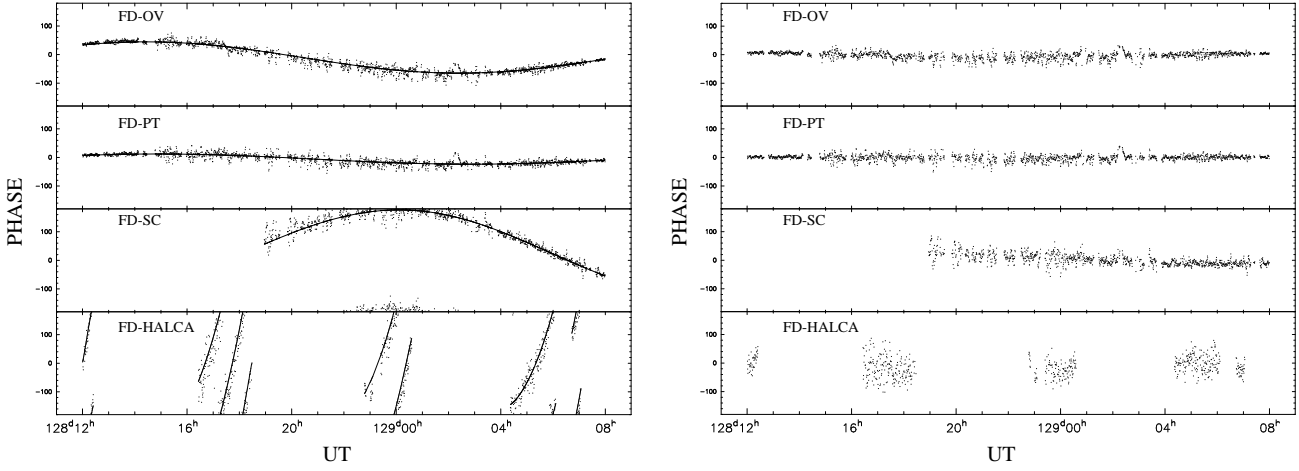}
\caption[]{Left panel: 5\,GHz phases of B\,1342+662 referenced to those 
of B\,1342+663 for a representative set of baselines. The continuous 
line corresponds to the phases derived from 
the phase-reference map of Fig. 4. Right panel: same as left panel 
after subtraction of the phases derived from the phase-reference map 
of Fig. 4 (continuous line in left panel). 
The symbols correspond to the following 
VLBA antennas: FD, Fort Davis; OV, Owens Valley; PT, Pie Town; 
SC, Saint Croix.}
\end{figure*}

For mapping purposes we transferred the data into the Caltech imaging 
program {\sc difmap} (Shepherd et al. 1995). We first performed several 
iterations of self-calibration in phase and amplitude using ground-based 
data alone; we proceeded 
further by adding data from HALCA to the ground-based data set. 
We used uniform weighting to obtain maps at maximum resolution (Fig. 2). 
It can be seen that using HALCA baselines, jet-like features are discernible 
towards the northeast in B\,1342+662 and towards the southwest for B\,1342+663, 
while at ground resolutions both sources appear unresolved.\\

\section{Results and Discussion}

\subsection{Phase-reference mapping with HALCA}

\noindent
Having direct detections of both sources for all baselines to HALCA allows us 
to calibrate the  goodness of the phase-reference technique. We performed a 
phase-reference analysis within {\sc aips} 
(e.g. Beasley \& Conway 1995). Although we have for both sources 
healthy detections, we chose B\,1342+663 (the brighter source of the pair)
to be the reference source. Therefore, the antenna
phase calibrations of B\,1342+663 (free from structure effects by using
a hybrid map of the source) were applied to the B\,1342+662 data.\\

\noindent
The resulting phases of B\,1342+662, as said, referenced to the 
position of B\,1342+663, can be unambiguously tracked
along the complete experiment (see Fig. 3; left panel). This is
hardly a surprise for ground-ground baselines, but extremely important for
baselines to HALCA: given two radio sources with a similar relative geometry 
to that of our pair, the uncertainties in HALCA's orbit
determination do not prevent the prediction of the phases of one source
from knowledge of the phases of the other. The phases in Fig. 3 
(left panel)
show a clear sinusoidal behavior, a result of the combination of the
contributions of
the structure of the target source, uncancelled effects of
correlator model imperfections (necessarily angular-dependent effects), and
error in the relative coordinates of the two sources.

The error in the relative coordinates of the sources is by far the 
largest contributor to the non-zero reference-phases. To show it, we obtained 
a phase-reference map of B\,1342+662
(Fig. 4) by Fourier inversion of its (phase-referenced) visibilities.
Both the offset with respect to the origin and, to a much lesser degree,
the source structure (B\,1342+662 is fairly compact) account
for most of the systematic effects in the phases (see
continuous line in Fig. 3; left panel). Once the
structure contribution and the contribution of the offset position of 
B\,1342+662 are removed from the phases,
the residual phases do not show significant
systematic effects (Fig. 3; {right panel}). However, {the root-mean-square
(rms) of these residuals phases is $\sim$90\degr\, on HALCA baselines, which} 
provide 
an estimate of the inaccuracies of the satellite orbit reconstruction: 
at 5\,GHz, and counting on the strong cancellation produced by the small
separation of our sources, a rms of $\sim$90\degr\, of the residual phases
corresponds to $\sim$10\,m uncertainties in HALCA's position.
Further, the dispersion of the phases of some ground-ground baselines 
(e.g., baseline FD-SC in Fig. 3) is similar to that of the
ground-space baselines, meaning that a significant 
part of the rms of the ground-space residual phases is due to ``conventional"
correlator uncertainties (e.g. {geometry of the ground-based array and/or} 
atmospheric effects over the ground
station). Accordingly, our 10\,m uncertainty estimate
appears as a fairly conservative upper bound to the uncertainty in
reconstructing HALCA's orbit. This conjecture can be verified, as 
shown below, from the
astrometry carried out using different phase-reference maps of B\,1342+662.\\

\subsection{Differential Astrometry}

\noindent
The measurement of the position of the peak-of-brightness of the phase-reference 
image in Fig. 4 yields the differential correction to the a priori positions 
used in the correlator model. The peak-of-brightness of B\,1342+662 is located 
at position $-0.562\pm0.016$ mas and $1.383\pm0.008$ mas in right ascension 
and declination, respectively, where the quoted errors correspond solely on the 
uncertainty in determining the maximum of the brightness distribution in the 
image, based on the signal-to-noise ratio and the interferometric beam 
size.\\

\noindent
To calibrate the contribution of the ground-space baselines in the
relative astrometry, we made two additional phase-reference maps of B\,1342+662,
namely, (1) a map using only ground-ground baselines (Fig 5; upper panel),
and (2) a map using only ground-space baselines (Fig. 5; lower panel).
Each of the  maps and offsets in Fig. 5 is to be compared with the image 
shown in 
Fig. 4 using all data. The ground-only phase-reference map of B\,1342+662 
shows, as expected, a point-like source, while 
the HALCA-only phase-reference map shows a clearer defined core-jet structure. 
The offsets of the three structures from the centers of the maps in Figs. 4 and 5 
are extraordinarily similar. The difference between the offsets measured in any two
maps of Figs. 4 and 5 is smaller than 70$\mu$as. In particular, the
difference in the relative position measured in the ground-only map
(Fig. 5 upper panel) with respect to that of the HALCA-only map (Fig. 5 lower panel) may be 
interpreted as a rough estimate of the 
uncertainties in HALCA's position. Such a difference is $\sim$60$\mu$as. 
Again, taking into account the separation of 0\fdg 08 of our sources,
we derive an estimate of the HALCA fractional baseline error of
$\sim$2.1$\times$10$^{-7}$. For an average projected baseline of 12\,000 kilometers
this error corresponds to an uncertainty in HALCA's position as small as 3\,m.
This result is in agreement with that provided by
Porcas et al. (2000) that estimated
an error of about 2-5\,m in HALCA's position for a pair of sources
separated by 14'.\\

\begin{figure}[htpb]
\vspace*{250pt}
\includegraphics{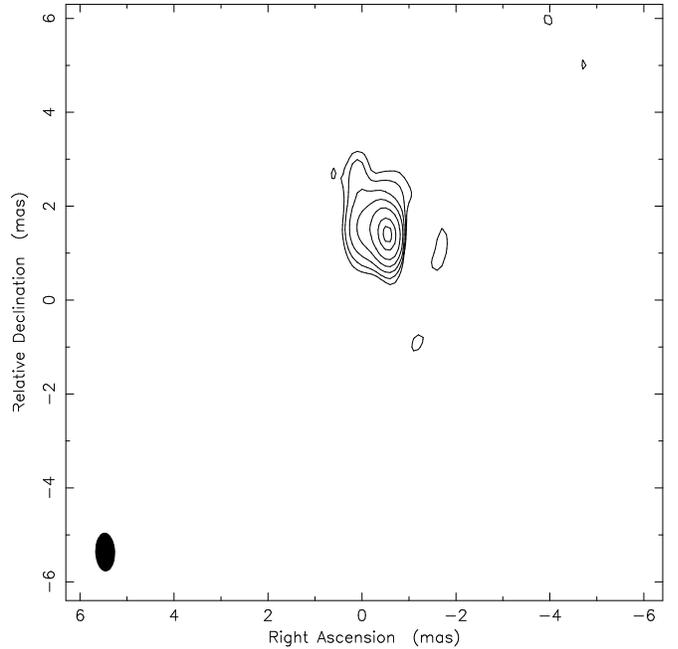}
\caption[]{5\,GHz map of B\,1342+662 phase-referenced to B\,1342+663. Contours 
are -2,2,4,8,16,32,64, and 90\% of the peak of brightness, 0.22 Jy/beam. 
The restoring beam (shown at the bottom left corner) is an elliptical Gaussian of
0.8$\times$0.4\,mas (P.A. 2\degr).}
\end{figure}

\begin{figure}[t]
\vspace*{400pt}
\includegraphics{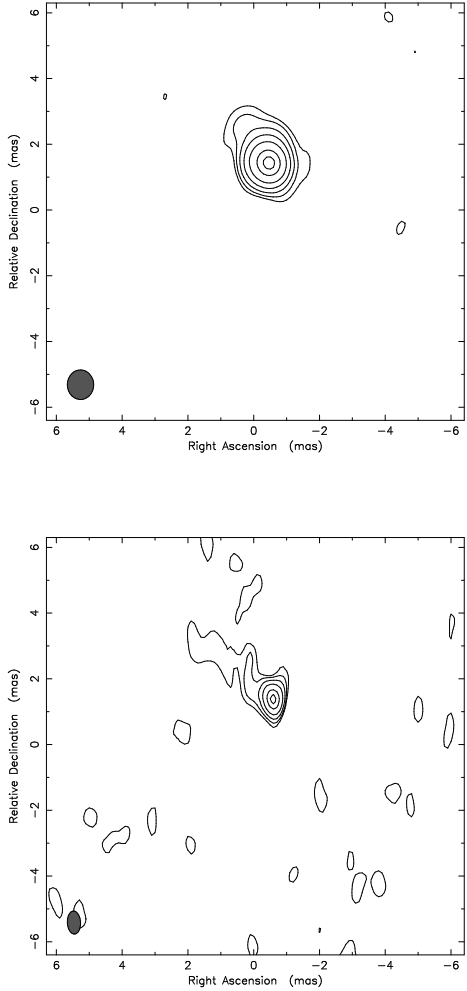}
\caption[]{Upper panel: 5\,GHz map of B\,1342+662 phase-referenced to
B\,1342+663 made with ground-ground data only. Contours are -2,2,4,8,16,32,64, and
90\% of the peak of brightness, 0.25 Jy/beam. The restoring beam (shown at
the bottom left corner) is an elliptical Gaussian of 0.9$\times$0.8\,mas
(P.A. 1\degr). Lower panel: same as upper panel but with
ground-space data only. Contours
are -4,4,8,16,32,64, and 90\% of the peak of brightness, 0.26 Jy/beam.
The restoring beam (shown at the bottom left corner) is an elliptical
Gaussian of 0.7$\times$0.4\,mas (P.A. 3\degr).}
\end{figure}

\noindent
Finally, the offset of the peak-of-brightness in the phase-reference map 
of B\,1342+662 (Fig. 4), and the a priori coordinates of the correlator model can be 
transformed to the J2000.0 {coordinates of the B\,1342+662 relative to 
those of B\,1342+663} at 5\,GHz:

\noindent
\begin{tabular}{lll}
$\Delta\alpha=$ & $-0^{h}\,0^{m}\,22\fs 720093$ & \, $\pm\,0\fs 000007$\\ 
$\Delta\delta=$ & $-0\degr\,3\arcmin\,45\farcs 898617$   & \,  $\pm\,0\farcs 000050$\\
\end{tabular}

\vspace*{0.2cm}
\noindent
where the quoted uncertainties have been enlarged to cover the
differences between the relative separation obtained by 
using the two maps made with ground-ground only and ground-space 
baselines only, respectively. We notice that the astrometric uncertainty
quoted above would not vary significantly from a sensitivity analysis of the 
different effects that contribute to the errors of the estimates of the 
relative coordinates. For pairs with such a small 
separation, most of this uncertainty 
comes from the reference point identification in the images 
(Marcaide et al. 1994; Rioja et al. 2000), whose contribution is already 
included in our quote.\\

\section{Conclusions}

\noindent
Our results show that phase-reference techniques can be applied to pairs of 
close radio sources using the HALCA satellite. The differential phase 
is unambiguously defined for ground-space baselines, with {a rms of the residual} 
phases of around 90\degr. Hence, radio sources with flux densities below HALCA's threshold 
detection can be observed if phase-referenced to nearby strong calibrators. 
We derive an upper bound of 10\,m for the uncertainties of HALCA's position. 
These uncertainties could be even smaller, as suggested by the relative astrometry 
of different sets of phase-reference maps
of our sources. With errors of this magnitude, HALCA is a useful 
tool for astrometry of close pairs of radio sources. 
The next generation of space-VLBI satellites will surely have a more accurate orbit 
determination (better than 10\,cm if based on GPS receivers aboard; 
Ulvestad 1999). This accurate positions over the entire orbit will facilitate 
the successful application of space-VLBI to astrometry and geodesy.

\acknowledgements{We thank Jim Ulvestad for a constructive refereeing 
of the paper and Richard Porcas for his valuable comments. This work has 
been supported by
the Spanish DGICYT grant PB96-0782. 
We gratefully acknowledge the VSOP Project, which is led by the 
Japanese Institute of Space and Astronomical Science in cooperation 
with many organizations and radio telescopes around the world.
Research at JPL is carried out under contract with the National 
Aeronautic and Space Administration. The National Radio Astronomy 
Observatory is operated by Associated Universities Inc., under a 
cooperative agreement with the National Science Foundation.}

\end{document}